\documentstyle[12pt,epsf]{article}
\def\fo{\hbox{{1}\kern-.25em\hbox{l}}}

\def\beq{\begin{equation}}
\def\eeq{\end{equation}}
\def\eq{\end{equation}}
\def\to{\rightarrow}

\def\bsg{\ifmmode B\to X_s\gamma\else $B\to X_s\gamma$\fi}
\def\bsll{\ifmmode B\to X_s\ell^+\ell^-\else $B\to X_s\ell^+\ell^-$\fi}
\def\bstt{\ifmmode B\to X_s\tau^+\tau^-\else $B\to X_s\tau^+\tau^-$\fi}
\def\shat{\ifmmode \hat{s}\else $\hat{s}$\fi}

\newcommand{\newc}{\newcommand}

\newc{\lcal}{\int {\cal L}dt}
 
\newc{\LSP}{{\chi^0_1}}
\newc{\stauR}{{\tilde \tau_R}}
\newc{\stau}{{\tilde \tau_1}}
\newc{\mstop}{m_{\tilde{t}}}
\newc{\mHpm}{m_{H^\pm}}
\newc{\gsim}{\lower.7ex\hbox{$\;\stackrel{\textstyle>}{\sim}\;$}}
\newc{\lsim}{\lower.7ex\hbox{$\;\stackrel{\textstyle<}{\sim}\;$}}
\newc{\ie}{{\it i.e.}}          
\newc{\etal}{{\it et al.}}
\newc{\eg}{{\it e.g.}}          
\newc{\kev}{\hbox{\rm\,keV}}            
\newc{\mev}{\hbox{\rm\,MeV}}            
\newc{\gev}{\hbox{\rm\,GeV}}            
\newc{\tev}{\hbox{\rm\,TeV}}
\newc{\xpb}{\hbox{\rm\, pb}}
\newc{\xfb}{\hbox{\rm\, fb}}

\newc{\mtop}{m_t}
\newc{\mbot}{m_b}
\newc{\mz}{m_Z}
\newc{\mw}{M_W}
\newc{\alphasmz}{\alpha_s(m_Z^2)}
\newc{\swsq}{\sin^2\theta_W}
\newc{\tw}{\tan\theta_W}
\newc{\cw}{\cos\theta_W}
\newc{\sw}{\sin\theta_W}
\newc{\BR}{\hbox{\rm BR}}
\newc{\zbb}{Z\to b\bar}
\newc{\Gb}{\Gamma (Z\to b\bar b)}
\newc{\Gh}{\Gamma (Z\to \hbox{\rm hadrons})}
\newc{\rbsm}{R_b^\hbox{\rm sm}}
\newc{\rbsusy}{R_b^\hbox{\rm susy}}
\newc{\drb}{\delta R_b}

\newc{\sgn}{\mbox{sgn}}
\newc{\tbeta}{\tan\beta}
\newc{\uL}{{\tilde u_L}}
\newc{\uR}{{\tilde u_R}}
\newc{\cL}{{\tilde c_L}}
\newc{\cR}{{\tilde c_R}}
\newc{\tL}{{\tilde t_L}}
\newc{\tR}{{\tilde t_R}}
\newc{\dL}{{\tilde d_L}}
\newc{\dR}{{\tilde d_R}}
\newc{\sL}{{\tilde s_L}}
\newc{\sR}{{\tilde s_R}}
\newc{\bL}{{\tilde b_L}}
\newc{\bR}{{\tilde b_R}}
\newc{\eL}{{\tilde e_L}}
\newc{\eR}{{\tilde e_R}}
\newc{\mhp}{m_{H^\pm}}
\newc{\mhalf}{m_{1/2}}
\newc{\emt}{{e/\mu /\tau}}

\newc{\lR}{\tilde{l}_R}
\newc{\lL}{\tilde{l}_L}
\newc{\nL}{\tilde{\nu}_L}
\newc{\na}{\chi^0_1}
\newc{\nb}{\chi^0_2}
\newc{\nc}{\chi^0_3}
\newc{\nd}{\chi^0_4}
\newc{\ca}{\chi^{\pm}_1}
\newc{\cb}{\chi^{\pm}_2}
\newc{\camp}{\chi^\mp_1}
\newc{\cbmp}{\chi^\mp_1}
\newc{\capos}{\chi^{+}_1}
\newc{\caneg}{\chi^{-}_1}
\newc{\phit}{\phi_t}
\newc{\phib}{\phi_b}
\newc{\phiew}{\phi_{ew}}
\newc{\htz}{h^0_t}
\newc{\hbz}{h^0_b}
\newc{\hewz}{h^0_{ew}}
\newc{\hsmz}{h^0_{sm}}
\newc{\huz}{h^0_u}
\newc{\hsusyz}{h^0_{susy}}

\def\EPC#1#2#3{Eur. Phys. J. C {\bf #1}, #3 (19#2)}
\def\NPB#1#2#3{Nucl. Phys. B {\bf #1}, #3 (19#2)}
\def\PLB#1#2#3{Phys. Lett. B {\bf #1}, #3 (19#2)}

\def\PRD#1#2#3{Phys. Rev. D {\bf #1}, #3 (19#2)}
\def\PRL#1#2#3{Phys. Rev. Lett. {\bf#1}, #3 (19#2)}

\def\beq{\begin{equation}}
\def\eeq{\end{equation}}
\def\bea{\begin{eqnarray}}
\def\eea{\end{eqnarray}}
\def\slashchar#1{\setbox0=\hbox{$#1$}           
   \dimen0=\wd0                                 
   \setbox1=\hbox{/} \dimen1=\wd1               
   \ifdim\dimen0>\dimen1                        
      \rlap{\hbox to \dimen0{\hfil/\hfil}}      
      #1                                        
   \else                                        
      \rlap{\hbox to \dimen1{\hfil$#1$\hfil}}   
      /                                         
   \fi}                                         %
\catcode`@=11
\long\def\@caption#1[#2]#3{\par\addcontentsline{\csname
  ext@#1\endcsname}{#1}{\protect\numberline{\csname
  the#1\endcsname}{\ignorespaces #2}}\begingroup
    \small
    \@parboxrestore
    \@makecaption{\csname fnum@#1\endcsname}{\ignorespaces #3}\par
  \endgroup}
\catcode`@=12

\def\jfig#1#2#3{
 \begin{figure}
 \centering
 \epsfysize=3.0in
 \hspace*{0in}
 \epsffile{#2}
 \caption{#3}
 \label{#1}
 \end{figure}}

\begin{document}

\begin{titlepage}

\begin{flushleft}
\end{flushleft}
\begin{flushright}
CERN-TH/98-308 \\
hep-ph/9809504\\
\end{flushright}

\vspace*{15.3cm}

\begin{flushleft}
CERN-TH/98-308 \\
September 1998
\end{flushleft}

\vspace*{-17.0cm}



\huge
\begin{center}
{\Large\bf
Supersymmetric dark matter \\
with a cosmological constant}
\end{center}

\large

\vspace{.15in}
\begin{center}

James D.~Wells

\vspace{.1in}
{\it CERN, Theory Division \\
 CH-1211 Geneva 23 \\}

\end{center}
 
 
\vspace{0.15in}
 
\begin{abstract}

Recent measurements of cosmological parameters from the microwave
background radiation, type Ia supernovae, and the age of globular clusters
help determine the relic matter density in the universe.
It is first shown with mild cosmological assumptions 
that the relic matter density satisfies
$\Omega_M h^2 < 0.6$ independent of the cosmological constant and
independent of the SNIa data.  Including the SNIa data, the constraint
becomes $\Omega_M h^2 < 0.35$.
This result is then applied to supersymmetric models 
motivated by generic features in supergravity mediated supersymmetry
breaking.  The result is an upper bound on gaugino masses
within reach of the LHC and a $1.5\tev$ lepton collider. Thus,
cosmological considerations are beginning to limit the supersymmetric
mass spectra in the experimentally verifiable range without recourse
to finetuning arguments, and without assuming a zero cosmological
constant.

\end{abstract}

\end{titlepage}

\baselineskip=18pt

\vfill
\eject


The fields of particle physics and cosmology overlap and enlighten
each other in many areas, including inflation, big bang nucleosynthesis,
cosmic rays, and dark matter.  The cosmology of dark matter, in particular,
has been an effective slayer of otherwise reasonable particle physics
models.  Often times, the amount of cold dark matter left over today is 
calculable in a theory~\cite{relic}, 
and may even yield a relic density too high to
be compatible with experiment.  In supersymmetry, this ``relic abundance
constraint'' has been a powerful one since it generally leads to
upper bounds on the mass of supersymmetric particles~\cite{drees93,kane94}.  
No other known
physics argument limits 
the mass of superpartners.  For this reason, dark matter
relic abundance has a unique role in supersymmetry.

In this letter the
recent cosmological measurements of the cosmic microwave background radiation
(CMB), age of the universe, Hubble constant, and supernova type Ia data
are combined to demonstrate that an upper bound exists
on $\Omega_M\equiv \rho_M /\rho_c$ (matter relic abundance), independent
of the cosmological constant.  These cosmological 
measurements greatly restrict supersymmetric parameter space.
One way to characterize the resulting
allowed supersymmetry parameter space is to show mass limits in
``generic supersymmetry''.  Generic
supersymmetry, as it will be defined in later paragraphs, merely implements
the expected hierarchy in soft supersymmetry breaking masses:  sleptons lighter
than squarks, $|\mu |\gg M_1$, etc.  In this case, the bino is the lightest
supersymmetric partner (LSP), and its relic abundance depends mostly 
on the mass of right-handed sleptons.


One way to begin a discussion of relic matter density in the universe
is to assume that the cosmological constant is zero.  Historically,
this has been an uncriticized assumption since one would naturally
assume that the cosmological constant is greater than
$m_{weak}^4$ or 0.
Since $m_{weak}^4$ is grossly incompatible with experiment
by many orders of magnitude, it appears that $0$ is the most tenable option.
However, recent experiments~\cite{perlmutter,highz} 
claim evidence for $\Omega_\Lambda\neq 0$, and recent theoretical 
work~\cite{nonzero}
has been entertaining once again the cosmological constant.
This opens the mind to a non-zero, but small, cosmological constant.

From a particle physics point of view, limits on
$\Omega_M h^2$ have traditionally been applied with the assumption that
$\Lambda =0$.  The $\Lambda =0$
assumption was so pervasive that it was not listed as a qualifier
for the $\Omega_M h^2$ bounds presented in the 1996 Particle Data Group 
Book~\cite{pdg96}.  The most
recent PDG book~\cite{pdg98}
rightly indicates that the listed bounds on $\Omega_M h^2$ are with
$\Lambda =0$. The most often cited limit on relic
matter density is $\Omega_M h^2 < 1$, which is derivable from the assumptions
that the universe is more than 10 billion years old, that
$\Lambda =0$, and that the Hubble
constant is greater than $40\, {\rm km\, s^{-1}\, Mpc^{-1}}$.  This constraint
has been widely applied in the particle physics community to place
restrictions on particle parameter space.  

If we relax the assumption that $\Lambda =0$, the age of the 
universe can be calculated by integrating the Friedmann equation with
appropriately scaled matter and vacuum energy densities,
\beq
t_{\rm age}=H^{-1}_0\int_0^1 da \sqrt{\frac{a}{(1-a)\Omega_M
 +(a^3-a)\Omega_\Lambda +a}}.
\eeq
The lower limit on $t_{\rm age}$ and $h$
{\em alone} can no longer put an upper limit on the relic
matter density in the universe.  It must be combined with another
cosmological observable.  
One such observable is the anisotropy of the cosmic microwave background
radiation.  Measurements of the lower multipole moment power spectrum
by COBE allow one to place bounds on the total energy 
density today~\cite{yamamoto96,white96}:
\beq
0.3 < \Omega_M+\Omega_\Lambda < 1.5 ~~~ {\rm (CMBR~constraint).}
\eeq
In the $\Omega_M - \Omega_\Lambda$ plane, this restriction is almost
perpindicular to the age of the universe constraint.  As we can see from
Fig.~\ref{summary}, if we utilize both constraints a maximum value of
$\Omega_M$ is derivable.  

Presently, the Hubble constant is known to be 
$H_0= 100 h \, {\rm km\, s^{-1}\, Mpc^{-1}}$, where~\cite{pdg98}
\beq
0.6 < h < 0.8 ~~~{\rm (Hubble~constant~range).}
\eeq
New measurements and data analysis~\cite{chaboyer97} 
indicate that the age of globular
clusters is $t_{\rm age}= 11.5\pm 1.3\, {\rm Gyr}$.  This can be used to
set a $95\%$ C.L. lower limit on the age of the universe of
$t_{\rm age}>9.5\, {\rm Gyr}$.
Therefore, 
the maximum area that the combined Hubble constant and age of the universe
measurements fill in the $\Omega_M - \Omega_\Lambda$ plane is between
the lines of $t_{\rm age}=9.5\, {\rm Gyr}$ with $h>0.6$ and approximately
$t_{\rm age}=15\, {\rm Gyr}$ with $h<0.8$.  In Fig.~\ref{summary} these
two lines are drawn and labelled, and the arrows indicate the area on the
plot allowed.

The maximum acceptable 
value of $\Omega_M$ is therefore at the cross-point
of the $t_{\rm age}=9.5\, {\rm Gyr}$ with $h>0.6$ line and
the $\Omega_M+\Omega_\Lambda < 1.5$ line.  Self-consistently
using $h=0.6$, one obtains the limit
\beq
\label{omatter}
\Omega_M < 1.7 ~~~{\rm and}~~~ 
\Omega_M h^2 < 0.6  ~~~{\rm (limit~on~matter~density).}
\eeq
It is not completely obvious that choosing the constraints with
$h=0.6$ yields the maximum value of $\Omega_M h^2$.  If we instead
chose a value of $h$ larger than $0.6$ and self-consistently checked
where the age of the universe constraint line met the CMBR line, we
would find a smaller value for the maximum $\Omega_M h^2$ allowed.  Therefore,
Eq.~\ref{omatter} is valid for all of allowed cosmological parameter
space. 
The result does not take into
account possible tensor fluctuations, etc., which if present, would alter
the allowed range~\cite{fits}.

The recent data analysis of type Ia 
supernovae~\cite{perlmutter,highz,white98} 
show evidence
for a cosmological constant.  Fig.~\ref{summary} plots
in the $\Omega_M - \Omega_\Lambda$ parameter plane the published results
of the Supernova Cosmology Project.  Estimated {\em maximum} systematic
errors have been included in the plot to make the allowed area 
conservatively large.  The results of the high-Z Supernova 
Search Team are similar.  In Fig.~\ref{summary} we see that the
supernova results are almost parallel in the plane to the age of
the universe constraint.  Therefore, the supernovae
data can effectively replace the direct application
of the universe's age  measurement in
relic matter density considerations.  As it stands, the
supernovae data combined with the CMBR implies
\beq
\Omega_M< 0.95 ~~~{\rm and}~~~ \Omega_M h^2 < 0.35~~~{\rm (limit~on}~\Omega_M~
{\rm from~SN1a~and~CMBR}),
\eeq
which is presently much more restrictive than the age of universe constraint
plus CMBR.

\jfig{summary}{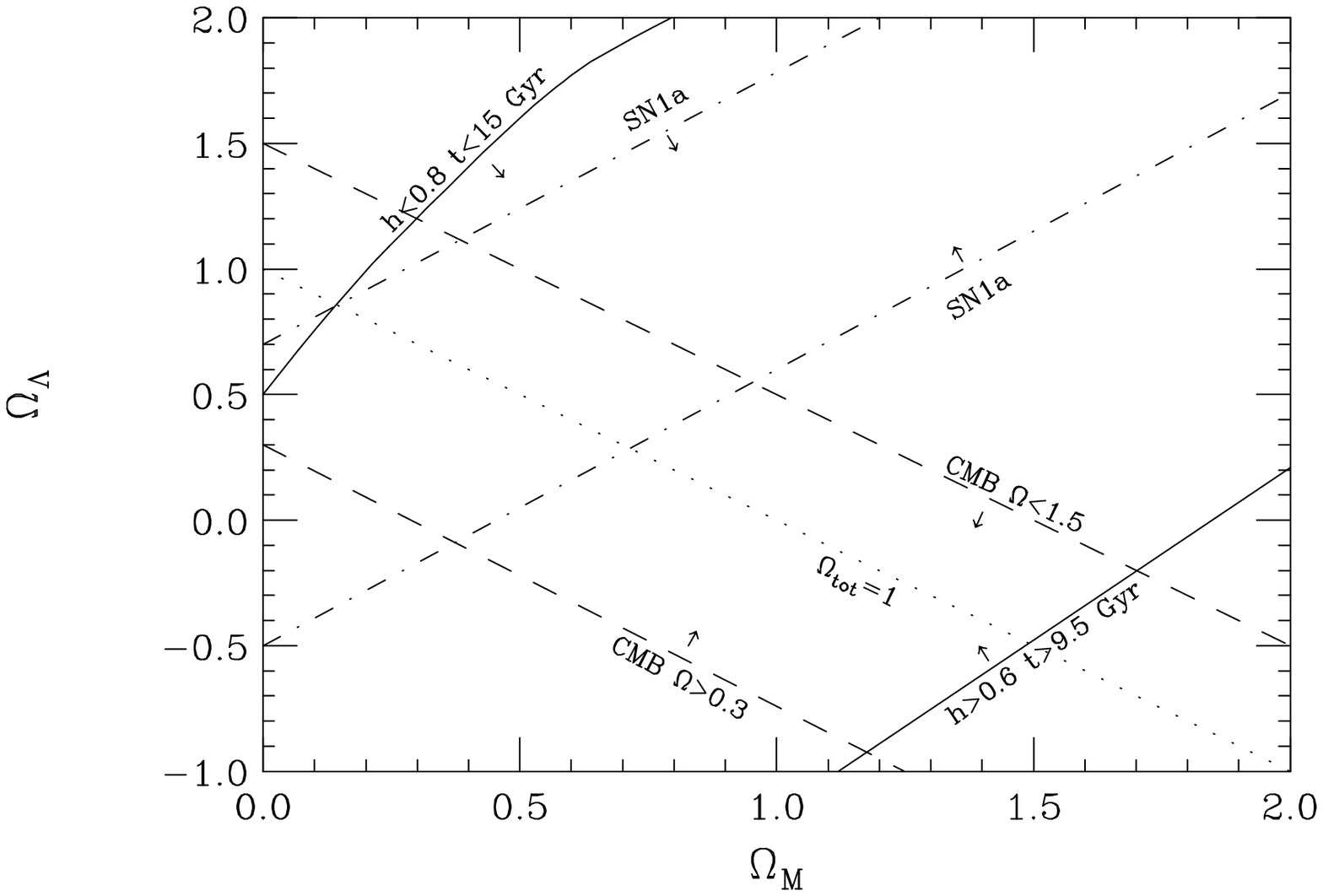}{The allowed area from the combination of the
Hubble constant measurement and the age of the universe measurement
are between the two solid lines.  The allowed region from SN1a data
is between the two dot-dashed lines.  And the allowed region from the
cosmic microwave background measurements are between the two dashed
lines.}


To apply these cosmological results to supersymmetry, a reasonable
model of supersymmetry should be defined that has a tractable
number of parameters.  To this end, one can identify
features of supersymmetry that are commonly manifested
in different approaches to model building.
One is that the soft-scalar masses are generation independent
and that the masses fall in a hierarchy defined
by the strength of their gauge interactions. This is true,
for example, in minimal 
supergravity models and $SO(10)$
grand unified models with a common mass for all scalars.
The hierarchy is generated by logarithms when the masses are renormalized
from their high scale values to their low scale values from gauge interactions.
It is therefore not unreasonable to assume 
a generic model of scalar superpartners
arranged according to each scalar's gauge interactions.  One should also
note that gauge-mediated models also demonstrate a hierarchy in scalar
masses according to the gauge interaction strength of each scalar.
Although the gravitino is generally the LSP in
gauge mediation models, it is nevertheless another example of how
specific model building usually implies a hierarchy
among the scalar states.

The important inference here from the generic supersymmetry model is
that $m_{\tilde l_R}$ is the lightest scalar, and that the gauginos
satisfy GUT relations ($M_i/\alpha_i={\rm const}$).  One additional
assumption is that the Higgsino mass term $\mu$ is sufficiently higher
than $M_1$ so that the lightest gaugino is almost pure 
bino~\cite{roszkowski91}. This is a
reasonable supposition, and is 
realized, for example, over most of parameter space in minimal 
supergravity~\cite{kane94}.

If nature is described by a supersymmetric spectrum roughly along the pattern 
described above, then the relic abundance will be 
calculable~\cite{susydm,drees93} from
only two parameters:  the masses of $m_{\tilde l_R}$ and $m_{\tilde B}$.
The formula is
\beq
\Omega_\chi h^2=\frac{(m^2_{\tilde l_R}+m^2_\chi)^4}{M^2\sqrt{N_F}
  m_\chi^2 (m^4_{\tilde l_R}+m^4_\chi)} 
\eeq
where $M \simeq 460\gev$ and $N_F$ is the number of degrees of freedom
at $\chi$-decoupling.

In Fig.~\ref{susy} the contours of $\Omega_\chi h^2$ are plotted in
the $m_{\tilde l_R} - m_{\tilde B}$ plane.
\jfig{susy}{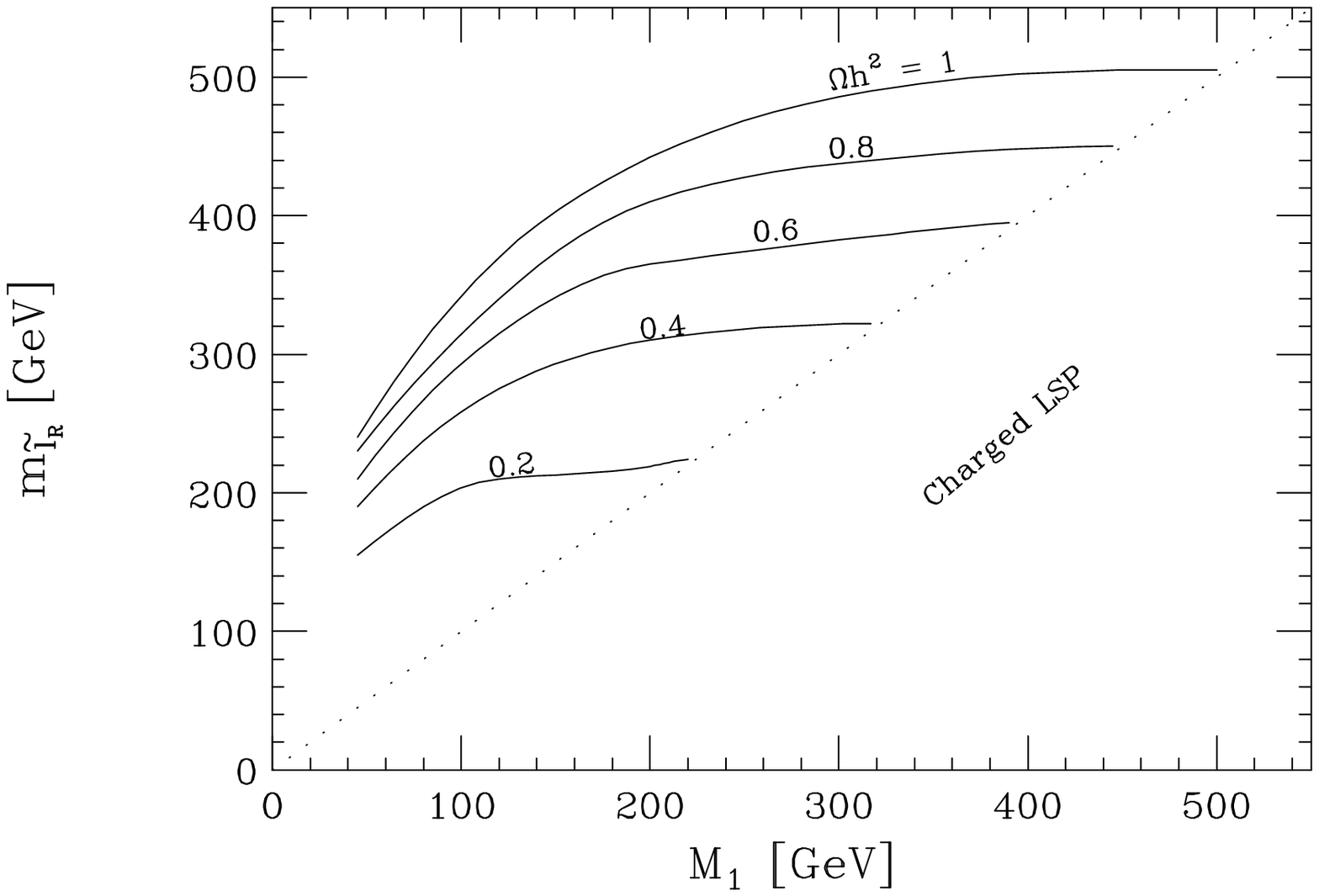}{Contours of constant $\Omega_\chi h^2$ in the
$m_{\tilde l_R}-m_\chi$ plane.  The old constraint $\Omega_\chi h^2 <1$
allowed bino (gluino) masses above $500\gev$ ($3\tev$), whereas
the new constraint $\Omega_\chi h^2 < 0.4$ allows bino (gluino)
masses only up to $300\gev$ ($2\tev$) at most, and therefore is
detectable at the LHC. The contours terminate at the left due to
chargino mass bounds from LEPII.}
The old constraint, $\Omega_\chi h^2 <1$, based on the age of the universe
and $\Lambda=0$ limits~\cite{pole}
$m_{\tilde l_R}\lsim 500\gev$.  The corresponding requirement 
of $m_{\tilde B}\lsim 500\gev$ is similar to the limit of $m_{1/2}\lsim 1\tev$
found in ref.~\cite{kane94} 
(with $m_t=175\gev$).  This is another indication that
the  model discussed above yields very similar results to
minimal supergravity, even though generic supersymmetry does not
specify the precise values of the squark masses, etc.  

The new constraint,
$\Omega_M h^2 <0.35$, based on the supernovae data and
the CMB radiation limits~\cite{pole} 
$m_{\tilde l_R}\lsim 300\gev$.  This is, of course,
a quantitative improvement over the old constraint.  However, this
is not just to say that a portion of parameter
space has been hacked off by this new constraint.  Rather, there is
a qualitative difference between a $500\gev$ and a $300\gev$ mass
limit on $m_{\tilde B}$.  In the first case, a $500\gev$ bino mass corresponds
to a gluino mass over $3\tev$. Gluinos over about $2-2.5\tev$ are not
likely to be detected at the LHC~\cite{baer96}.  
However, the $300\gev$ mass limit
on $m_{\tilde B}$ corresponds to a mass limit of the gluino of
about $2\tev$.  This is detectable at the LHC~\cite{baer96}, and
also a $1.5\tev$ center-of-mass energy lepton collider.  Of course,
these are upper bounds on the mass within the generic supersymmetry
model, and the actual mass may be much lower.
The conclusion here
is that {\em general expectations of supersymmetry along with
cosmology requirements produce a spectrum of supersymmetric states 
visible at the LHC and a $1.5\tev$ lepton collider.}  
Finetuning arguments are not needed in
this framework to make the
prediction that these colliders can find superpartners.

The correlation made above between relic abundance of the LSP and detectability
at the LHC and a $1.5\tev$ lepton collider
is straightforward.  This is because the LSP mass 
correlates with all gaugino masses,
and detectability at the colliders is closely linked to gaugino
production processes.  In other words, a minimum cross-section
of non-SM signatures is guaranteed just from gauginos
whose properties are known from relic abundance
considerations and a few assumptions about the supersymmetry mass 
spectrum enumerated above.
Correlations between dark matter and lower limits from LEP2 collider searches
also yield interesting restrictions on parameters space~\cite{ellis}.

Correlating dark matter relic abundance in generic supersymmetry
with direct searches for dark matter is not so straightforward.
For example, a small higgsino component to the LSP will not change
the relic abundance calculation in a noteworthy way; however, a small
higgsino component can change the cryogenic direct detection rates
of LSP-nucleon scattering substantially~\cite{griest,diehl95}.  
Therefore, the direct
detection method depends very sensitively on the $M_1/\mu$ ratio, whereas
the relic abundance does not as long as $|\mu|\gsim 2 M_1$.

Searches for supersymmetric dark matter from annihilations of LSPs in the
galactic halo are also not easily correlated with the relic abundance
in generic supersymmetry.  The reason is because LSP virial velocity
in the galactic halo is only a few hundred kilometers per second, and thus
non-relativistic.  All annihilations proceed through a helicity
suppressed $S$-wave and want
to terminate in a heavy quark, such as the $b$ quark, rather than
a lepton.  Annihilations in the early universe which dictate the relic
abundance are done in a sufficiently relativistic regime such that
$P$-wave annihilations, which are
not helicity suppressed, can dominate.  Final state
leptons from $\chi\chi$ annihilations through a $t$-channel slepton
are of primary importance.  Therefore, since a small higgsino content or
heavy squarks can mediate significant 
$S$-wave annihilations in the galactic halo 
(non-relativistic limit), they are in principle
totally uncorrelated with the relic
abundance.  For this reason, no firm predictions can be made about these processes,
even if one had full knowledge of the density profile of dark
matter in the galactic halo.

Although colliders cannot easily tell if a particle lives for more
than a few meters of $c\tau$, they are effective probes of dark 
matter~\cite{diehl95,baer98}.
As was discussed above, collider physics observables are cleanly correlated
with relic abundance in generic supersymmetry, whereas all dark matter
specific experiments are not necessarily correlated.  

In conclusion, $\Omega_M h^2 <0.35$ is perhaps a
more appropriate constraint to apply to particle physics models than
the old $\Omega_M h^2 <1$ constraint.  This new constraint relies on
the lower bound of the age of the universe and Hubble constant, 
SNIa data, and the upper bound of total energy density of the universe from
microwave background measurements.  It does not depend on $\Lambda=0$.
Applying this constraint to generic supersymmetry -- 
a model of supersymmetry based on the gauge coupling hierarchy -- one
finds that the allowed parameter space is within reach of the LHC,
and a $1.5\tev$ lepton collider.
Other approaches to supersymmetry breaking which have
family dependent masses, for example, also can be correlated with
relic abundance constraints~\cite{berezinskii96,nath97,gherghetta} and
limits can be obtained in these frameworks.
In short, recent and forthcoming
cosmological parameter measurements are significantly restricting the
parameter space of any model of supersymmetry breaking which has
a heavy, stable LSP.

\bigskip
\noindent
{\it Acknowledgements:} I thank T.~Gherghetta, S.~Martin and A.~Riotto
for helpful conversations.



\begin{thebibliography}{20}

\bibitem{relic}
B.~Lee and S.~Weinberg, \PRL{39}{77}{165};
M.~Vysotskii, A.~Dolgov, and Ya.~Zeldovich, Pisma Zh. Eksp. Teor. Fiz.
{\bf 26}, 200 (1977);
P.~Hut, \PLB{69}{77}{85}.

\bibitem{drees93}
M.~Drees and M.~Nojiri, \PRD{47}{93}{376}.

\bibitem{kane94}
G.L.~Kane, C.~Kolda, L.~Roszkowski, and J.D.~Wells, \PRD{49}{94}{6173}.

\bibitem{perlmutter}
S.~Perlmutter et al., Nature {\bf 391}, 51 (1998).

\bibitem{highz}
A.G.~Riess et al., astro-ph/9805201.

\bibitem{nonzero}
See, as one example, N.~Turok and S.W.~Hawking, \PLB{432}{98}{271}.

\bibitem{pdg96}
Particle Data Group (R.M.~Barnett et al.),
{\it Review of Particle Physics}, \PRD{54}{96}{1}.

\bibitem{pdg98}
Particle Data Group (C.~Caso et al.), 
{\it Review of Particle Physics}, \EPC{3}{98}{1}.

\bibitem{yamamoto96}
K.~Yamamoto and E.~Bunn, Astrophys. J. {\bf 464}, 8 (1996).

\bibitem{white96}
M.~White and D.~Scott, Astrophys. J. {\bf 459}, 415 (1996).

\bibitem{chaboyer97}
B.~Chaboyer, P.~Demarque, P.~Kernan, and L.~Krauss, 
Astrophys. J.~{\bf 494}, 96 (1998).

\bibitem{fits}
C.~Lineweaver, astro-ph/9805326; M.~Tegmark, astro-ph/9809201.

\bibitem{white98}
M.~White, astro-ph/9802295.

\bibitem{roszkowski91}
L.~Roszkowski, \PLB{262}{91}{59}.


\bibitem{susydm}
H.~Goldberg, \PLB{50}{83}{1419};
M.~Srednicki, R.~Watkins, and K.~Olive, \NPB{310}{88}{693};
P.~Gondolo and G.~Gelmini, \NPB{360}{91}{145};
G.~Jungman, M.~Kamionkowski, K.~Griest, Phys. Rep. {\bf 267}, 195 (1996).

\bibitem{pole}
Limits are valid as long as $m_\chi$ is not equal to approximately
one-half the mass of a Higgs boson such that $\chi\chi$ annihilations
could proceed through a Higgs boson resonance, thereby lowering the relic
abundance.

\bibitem{baer96}
H.~Baer, C.-H.~Chen, F.~Paige, and X.~Tata, \PRD{53}{96}{6241} and
\PRD{52}{95}{2746}.

\bibitem{ellis}
J.E.~Ellis, T.~Falk, K.A.~Olive, and M.~Schmitt, \PLB{413}{97}{355}.

\bibitem{griest}
K.~Griest, \PRD{38}{88}{2357}, {\it erratum ibid.} D {\bf 39}, 3802 (1989).

\bibitem{diehl95}
E.~Diehl, G.L.~Kane, C.~Kolda, and J.D.~Wells, \PRD{52}{95}{4223}.

\bibitem{baer98}
H.~Baer and M.~Brhlik, \PRD{57}{98}{567}.


\bibitem{berezinskii96}
V.~Berezinskii, A.~Bottino, J.~Ellis, N.~Fornengo, G.~Mignola,
and S.~Scopel, Astropart. Phys. {\bf 5}, 1 (1996).

\bibitem{nath97}
P.~Nath and R.~Arnowitt, \PRD{56}{97}{2820}.

\bibitem{gherghetta}
T.~Gherghetta, A.~Riotto, and L.~Roszkowski, hep-ph/9804365.


\end{thebibliography}
\end{document}